\documentclass[aps,prb,superscriptaddress,twocolumn,floatfix]{revtex4-1}
\usepackage{graphicx}
\usepackage{dcolumn}
\usepackage{bm}
\usepackage{stmaryrd}
\usepackage{latexsym}
\usepackage{amssymb}
\usepackage{amsfonts}
\usepackage{amsmath}
\usepackage[breaklinks=true,colorlinks,citecolor=blue,linkcolor=blue,urlcolor=blue]{hyperref}

\begin{document}
\title{Quantum capacitance and Landau parameters of massless Dirac fermions in graphene}

\author{Reza Asgari}
\affiliation{School of Physics, Institute for Research in Fundamental Sciences (IPM), Tehran 19395-5531, Iran}
\author{Mikhail I. Katsnelson}
\affiliation{Radboud University Nijmegen, Institute for Molecules and Materials, NL-6525 AJ Nijmegen, The Netherlands}
\author{Marco Polini}
\affiliation{NEST, Istituto Nanoscienze-CNR and Scuola Normale Superiore, I-56126 Pisa, Italy}
\begin{abstract}
We present extensive numerical results for the thermodynamic density of states (i.e.~quantum capacitance) of a two-dimensional massless Dirac fermion fluid in a doped graphene sheet. In particular, by employing the random phase approximation, we quantify the impact of screening exerted by a metal gate located nearby a graphene flake. Finally, we calculate 
the spin- and circularly-symmetric Landau parameter, which can be experimentally extracted from independent measurements on the same setup of the quantum capacitance and quasiparticle velocity.
\end{abstract}

\maketitle

\section{Introduction}
In a series of seminal papers~\cite{Landau}, Landau formulated a very elegant macroscopic theory of normal Fermi liquids~\cite{Nozieres,Platzman_and_Wolff,shankar_rmp_1994,Giuliani_and_Vignale}. In his theory, the response to external perturbations of a system of interacting fermions (whose ground state is continuously connect to the ground state of the free Fermi gas) can be written in terms of a small set of dimensionless parameters known as ``Landau parameters". 

``Quasiparticles'', i.e.~dressed electrons, are the key players in a normal Fermi liquid. They move with a renormalized Fermi velocity $v^\star_{\rm F}$ and interact through the so-called Landau interaction function $f_{\sigma\sigma'}(\cos(\theta))$, where $\sigma$ and $\sigma'$ denote spin labels and $\theta$ is the angle between two wave vectors ${\bm k}$ and ${\bm k}'$ lying on the Fermi circle, i.e.~$|{\bm k}| = |{\bm k}'| = k_{\rm F}$, where $k_{\rm F}$ is the Fermi wave number. Interactions between two quasiparticles can be conveniently expanded in terms of dimensionless quantities $F^{{\rm s}, {\rm a}}_\ell$, where ``s" (``a") refers to the spin symmetric (antisymmetric) channel and $\ell = 0, 1, 2, \dots$ to the angular momentum channel. The quantities $F^{{\rm s}, {\rm a}}_\ell$ are the so-called Landau parameters.

The response of a system of interacting fermions to a perturbation that couples to circularly symmetric deformations of the 2D Fermi surface is controlled by the $\ell =0$ channel only. A well-known example is that of the compressibility $K$, which can be expressed as~\cite{Nozieres,Platzman_and_Wolff,shankar_rmp_1994,Giuliani_and_Vignale}
\begin{equation}\label{eq:Fermiliquid}
\frac{K}{K_{0}} = \frac{v_{\rm F}/v^\star_{\rm F}}{1+ F^{\rm s}_0}~.
\end{equation}
Microscopically, the compressibility $K$ is given by the following thermodynamic derivative: $n^2 K = \partial n/\partial \mu$. Here $\mu$ is the chemical potential, 
\begin{equation}\label{eq:mudef}
\mu = \frac{\partial [n \varepsilon(n)]}{\partial n}~, 
\end{equation}
with $\varepsilon(n)$ the ground-state energy per particle of the system of interacting fermions and $n$ is the density. Finally, $K_0$ is the compressibility of the non-interacting system. The compressibility is therefore a quantity of pivotal importance, since it is directly related to the ``equation of state'' $\varepsilon(n)$, carrying precious information about exchange and correlation contributions to the ground state energy per particle of the interacting system. 

In a two-dimensional (2D) parabolic-band electron gas in an ordinary semiconductor quantum well, exchange tends to enhance the charge response~\cite{Giuliani_and_Vignale} driving a change of sign of $\partial \mu/\partial n$ at a density $n^\star  = 2 a^{-2}_{\rm B}/\pi^3$, where $a_{\rm B} = \hbar^2 \epsilon/(m_{\rm b}e^2)$ is the material Bohr radius with $\epsilon$ a suitable dielectric constant and $m_{\rm b}$ the solid-state mass---for example, for GaAs $\epsilon \sim 13$ and $m_{\rm b} = 0.067~m_{\rm e}$, where $m_{\rm e}$ is the bare electron mass in vacuum. At the same density $K/K_0$ diverges. The sign change of the charge response of the 2D electron gas in a GaAs quantum well was measured for example by Eisenstein and collaborators~\cite{eisenstein_prl_1992,eisenstein_prb_1994} and the measured compressibility was found to be in excellent quantitative agreement with a simple Hartree-Fock estimate that takes into account the non-zero thickness of the GaAs quantum well.

The ability to isolate 2D ``atomic crystals''---like graphene and its derivatives---though micromechanical cleavage of their three-dimensional parent materials~\cite{kostya_pnas_2005,geim_naturemater_2007,Katsnelsonbook} has offered us an entirely new class of 2D electron systems with a number of exotic many-body properties~\cite{kotov_rmp_2012}.
In particular, the thermodynamic density-of-states (TDOS) $\partial n/\partial \mu$ of few-layer graphene sheets has been the subject of a number of a large number of experimental studies~\cite{martin_naturephys_2008,xia_naturenano_2009,xia_apl_2010,ponomarenko_prl_2010,henriksen_prb_2010,weitz_science_2010,young_prb_2012,yu_pnas_2013}. 

Martin {\it et al.}~\cite{martin_naturephys_2008} were the first to present experimental data for single-layer graphene, which were obtained through the use of a scanning single-electron transistor. Data in this work were rather noisy due to the high level of disorder of the used samples, which were exfoliated graphene sheets deposited on ${\rm SiO}_2$.  The main conclusion of this work was that the measured TDOS could be fit by a simple non-interacting formula for 2D massless Dirac fermions (MDFs),
\begin{equation}\label{eq:noninteracting}
n^2 K_0 = \left.\frac{\partial n}{\partial \mu}\right|_0 = \frac{2 \varepsilon_{\rm F}}{\pi \hbar^2 v^2_{\rm F}}~, 
\end{equation}
where $\varepsilon_{\rm F} = \hbar v_{\rm F} k_{\rm F}$ is the Fermi energy, $v_{\rm F}$ is the density-independent Fermi velocity, and $k_{\rm F} = \sqrt{\pi |n|}$ is the Fermi wave number. The absolute value in the previous expression ensures the applicability of Eq.~(\ref{eq:noninteracting}) to both electron- and hole-doped samples. The only quantity that can be used as a fitting parameter in Eq.~(\ref{eq:noninteracting}) is the Fermi velocity $v_{\rm F}$. By fitting the experimental data, the authors of Ref.~\onlinecite{martin_naturephys_2008} obtained the value $v^\star_{\rm F} =  1.1 \times 10^{6}~{\rm m}/{\rm s}$, which is larger than the bare value of the Fermi velocity given by tight-binding theory~\cite{castroneto_rmp_2009} or by density-functional theory at the level of the local-density approximation (LDA)~\cite{yang_prl_2009,schilfgaarde_prb_2011,attaccalite_PSSB_2009}.  The Fermi velocity enhancement is a well understood phenomenon stemming from electron-electron interactions~\cite{gonzalez_prb_1999,polini_ssc_2007,borghi_ssc_2009,elias_naturephys_2011} and is not captured by LDA~\cite{polini_ssc_2007}.

More recently, the TDOS of the 2D MDF liquid has been measured~\cite{yu_pnas_2013} in high-quality graphene sheets encapsulated in hexagonal Boron Nitride (hBN). The authors of Ref.~\onlinecite{yu_pnas_2013} adopted a fitting procedure similar to that used in Ref.~\onlinecite{martin_naturephys_2008}. The quality of the used samples and the experimental accuracy was so high, however, that they were able to unravel a non-trivial dependence of the quasiparticle velocity $v^\star_{\rm F}$, i.e.~the fitting parameter, on carrier density $n$. In particular, they found a logarithmic increase of $v^\star_{\rm F}$ upon lowering the carrier density towards the charge neutrality point. These results are in agreement with earlier results by Elias {\it et al.}~\cite{elias_naturephys_2011} for the quasiparticle velocity enhancement measured from the temperature dependence of the amplitude of the Shubnikov-de Haas oscillations in a weak magnetic field and in a suspended sample.

Theoretically, several calculations of the TDOS of the 2D MDF fluid in doped few-layer graphene sheets have appeared in the literature~\cite{barlas_prl_2007,hwang_prl_2007,abergel_prb_2009,borghi_prb_2010,abergel_prb_2011}. 
In particular, Barlas~{\it et al.}~\cite{barlas_prl_2007} pointed out that exchange interactions in a 2D MDF fluid tend to {\it suppress} the charge response, rather then enhancing it. They presented numerical results for the ratio $K/K_0$ calculated within the random phase approximation (RPA), showing clearly that $K/K_0<1$ in a 2D MDF fluid. RPA correlations were shown~\cite{barlas_prl_2007} not to be strong enough to counteract exchange interactions.

The aim of this Article is twofold. We first present numerical results based on the RPA for the TDOS of a doped graphene sheet in the presence of a nearby metal gate. We  demonstrate that the role of the metal gate is completely negligible for graphene sheets encapsulated between two insulators with equal dielectric constants (such as in the case of Ref.~\onlinecite{yu_pnas_2013}) and propose experiments where the role of screening exerted by the metal gate is predicted to play a much more important role. We then present a microscopic theory that quantifies the {\it difference} between $K/K_0$ and the quasiparticle velocity enhancement $v^\star_{\rm F}/v_{\rm F}$, highlighting the role of the spin- and circularly-symmetric Landau parameter $F^{\rm s}_0$. We propose an experiment that allows to measure $F^{\rm s}_0$, which will greatly help to clarify the role of vertex corrections in the many-body physics of doped graphene sheets. 

Our Article is organized as following. In Sect.~\ref{sect:model} we present our electrostatic model to deal with the presence of a metal gate close to a graphene sheet and we briefly summarize the theoretical approach we have used to calculate separately $K/K_0$ and $v^\star_{\rm F}/v_{\rm F}$. We report our main numerical results in Sect.~\ref{sect:numericalresults}, while our conclusions are reported in Sect.~\ref{sect:conclusions}.

\section{Modelling of the metal gate, basic physical parameters, and theoretical approach}
\label{sect:model}

\begin{figure}[t]
\begin{center}
\tabcolsep=0cm
\begin{tabular}{c}
\includegraphics[width=1.0\linewidth]{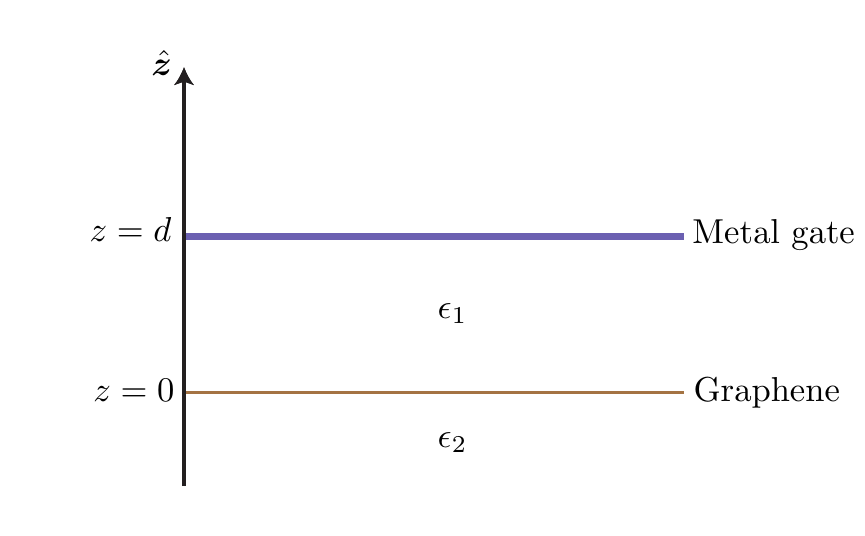}
\end{tabular}
\caption{(Color online) A graphene sheet is located at $z = 0$. A metal gate is located at $z = d$. The region of space $0 < z < d$ is filled with an insulator of thickness $d$ with dielectric constant $\epsilon_1$. The region of space $z<0$ is filled with an insulator with dielectric constant $\epsilon_2$. In the experiments of Ref.~\onlinecite{yu_pnas_2013} $\epsilon_1 = \epsilon_2 = 4.5$ corresponding to a graphene sheet encapsulated in hBN. In this work we carry out calculations also for the general case $\epsilon_1 \neq \epsilon_2$.\label{fig:one}}
\end{center}
\end{figure}

We focus on the setup depicted in Fig.~\ref{fig:one}, which exemplifies the one employed in recent experimental work~\cite{yu_pnas_2013}. The setup is composed by a metal gate located at a distance $d$ from a graphene sheet, which is encapsulated between two insulators with dielectric constants $\epsilon_1$ and $\epsilon_2$. In Ref.~\onlinecite{yu_pnas_2013} $\epsilon_1 = \epsilon_2 = 4.5$ since the graphene sheet was there encapsulated in hBN. Below we present numerical results for this case but also for $\epsilon_1 \neq \epsilon_2$.

Treating the metal gate as a perfect conductor, one can derive from elementary electrostatics the following 
effective interaction~\cite{tomadin_prb_2013,carrega_njp_2012} 
between two electrons bound to the graphene sheet:
\begin{equation}\label{eq:channelpotential}
V_q= \frac{2 \pi e^2}{{\bar \epsilon} q}\frac{1 - \exp{(-2 q d)}}{\displaystyle 1 + \frac{\epsilon_{1} - \epsilon_{2}}{\epsilon_{1} + \epsilon_{2}} \exp(-2 q d)} \equiv v_q{\cal F}(qd)~,
\end{equation}
where ${\bar \epsilon} \equiv (\epsilon_{1} + \epsilon_{2})/2$ and
\begin{equation}
{\cal F}(x) \equiv \frac{1 - \exp{(-2 x)}}{\displaystyle 1 + \frac{\epsilon_{1} - \epsilon_{2}}{\epsilon_{1} + \epsilon_{2}} \exp(-2 x)}~.
\end{equation}
Note that $V_q$ reduces to the 2D Fourier transform $v_q \equiv 2\pi e^2/({\bar \epsilon} q)$ of the usual bare Coulomb interaction in the limit $d \to \infty$. 

In a setup like the one depicted in Fig.~\ref{fig:one}, one can accurately measure the capacitance $C$, which can be easily shown~\cite{eisenstein_prl_1992,eisenstein_prb_1994,yu_pnas_2013} to contain two contributions ``in series'', i.e.
\begin{equation}\label{eq:series}
\frac{1}{C} \equiv \frac{1}{C_{\rm c}} + \frac{1}{C_{\rm q}}~,
\end{equation}
where $C_{\rm c}  =  S \epsilon_1/(4\pi d)$ is the classical capacitance and $C_{\rm q} \equiv Se^2 \partial n/\partial \mu$ is the so-called quantum capacitance. Here $S$ is the 2D electron system area.

In passing, we note that the {\it Hartree} contribution to the capacitance
\begin{equation}\label{eq:Hartree}
\frac{1}{C_{\rm H}}  \equiv \frac{1}{Se^2}\lim_{q \to 0} V_{q} =  \frac{1}{C_{\rm c}}
\end{equation}
coincides with the classical capacitance.

\subsection{Microscopic theory of the TDOS}

With the effective interaction in Eq.~(\ref{eq:channelpotential}) we can calculate the ground-state energy $\varepsilon(n)$ of the 2D MDF fluid in the graphene flake by employing the well-known~\cite{Giuliani_and_Vignale} fluctuation-dissipation and Hellman-Feynman theorems. This approach has been used by Barlas {\it et al.}~\cite{barlas_prl_2007} to calculate the TDOS of a 2D system of MDFs in the {\it absence} of a metal gate, i.e.~for $V_q \to 2\pi e^2/({\bar \epsilon} q)$ in Eq.~(\ref{eq:channelpotential}).

When the 2D MDF model is used to describe a doped graphene sheet, the quantity $\varepsilon(n)$ needs to be regularized~\cite{barlas_prl_2007} by subtracting off the (infinite) ground-state energy of the filled sea of negative energy states that the unbounded linear dispersion allows. Following Ref.~\onlinecite{barlas_prl_2007}, we therefore introduce the ground-state energy per excess electron (hole) $\delta \varepsilon(n)$, which is defined as the difference between $\varepsilon(n)$ and the ground-state energy of the charge neutral system. 

We find that $\delta \varepsilon(n) \equiv \delta \varepsilon_{\rm x}(n) + \delta \varepsilon_{\rm c}(n)$, where
\begin{equation}\label{eq:exchange}
\delta \varepsilon_{\rm x}(n) = -\frac{\hbar}{2\pi n}\int \frac{d^2 {\bm q}}{(2\pi)^2}~V_q \int_0^{+\infty}
d\Omega~\delta \chi_0(q,i\Omega)
\end{equation}
is the exchange contribution (i.e.~the first order contribution in powers of the electron-electron interaction potential $V_q$) and
\begin{widetext}
\begin{equation}\label{eq:correlation}
\delta \varepsilon_{\rm c}(n) = \frac{\hbar}{2\pi n}\int \frac{d^2 {\bm q}}{(2\pi)^2}\int_0^{+\infty}d\Omega\left\{V_q \delta \chi_0(q,i\Omega)+
\ln{\left[\frac{1-V_q\chi_0(q,i\Omega)}{1-V_q\left.\chi_0(q,i\Omega)\right|_{\varepsilon_{\rm F}=0}}\right]}\right\}
\end{equation}
\end{widetext}
is the RPA correlation contribution. Here
\begin{equation}
\delta \chi_0(q,i\Omega)  = \chi_0(q,i\Omega) - \left.\chi_0(q,i\Omega)\right|_{\varepsilon_{\rm F}=0}~,
\end{equation}
where $\chi_0(q,i\Omega)$ is the well-known~\cite{barlas_prl_2007}  non-interacting density-density response function of a system of 2D MDFs on the imaginary frequency axis and at finite density $n$, while
\begin{equation}
\left.\chi_0(q,i\Omega)\right|_{\varepsilon_{\rm F}=0} = -\frac{N_{\rm f} q^2}{16\hbar}\frac{1}{\sqrt{\Omega^2 + v^2_{\rm F} q^2 }}
\end{equation}
is the same quantity for an undoped system. Here, $N_{\rm f} = 4$ is the number of fermion flavors in graphene (spin and valley degrees of freedom). 

The integrals over $\Omega$ in Eqs.~(\ref{eq:exchange})-(\ref{eq:correlation}) are finite while the integrals over $q$ have logarithmic ultraviolet
divergences.  As explained in Ref.~\onlinecite{barlas_prl_2007}, these divergences are physical and follow from the interaction between
electrons near the Fermi energy and electrons very far from the Fermi
energy.  We introduce an ultraviolet cutoff for the wavevector integrals,
\begin{equation}\label{eq:cutoff}
k_{\rm c}  = \sqrt{\frac{2\pi}{{\cal A}_0}}~,
\end{equation}
where ${\cal A}_0=3\sqrt{3} a^2_0/2 \simeq 0.052~{\rm nm}^2$ is the area of the unit cell in the honeycomb lattice, $a_0 \simeq 1.42$~\AA~being the carbon-carbon distance. The 2D MDF model is useful for carrier densities such that $k_{\rm F} \ll k_{\rm c}$.

The TDOS can be easily calculated from Eq.~(\ref{eq:mudef}) with $\varepsilon(n) \to \delta \varepsilon(n)$. Note that the regularization scheme we have employed, i.e.~the definition of $\delta \varepsilon(n)$, does not affect the dependence of $\mu$ on $n$ since Eq.~(\ref{eq:mudef}) is sensitive only to changes of the ground-state energy with excess electron or hole density and not to the absolute magnitude of the ground-state energy.

\subsection{Microscopic theory of the quasiparticle velocity}

We now turn to summarize the microscopic theory~\cite{polini_ssc_2007} we have used to calculate the renormalized quasiparticle velocity $v^\star_{\rm F}$. We start from a microscopic expression for the quasiparticle Matsubara self-energy $\Sigma_s({\bm k},i\omega_n)$ in which this quantity is expanded to first order in the dynamically screened 
Coulomb interaction $W$ (setting $\hbar=1$):
\begin{widetext}
\begin{equation}\label{eq:sigma_rpa}
\Sigma_s({\bm k},i\omega_n)=-\frac{1}{\beta}\sum_{s' = \pm 1}
\int \frac{d^2{\bm q}}{(2\pi)^2}\sum_{m=-\infty}^{+\infty}
W(q,i\Omega_m)\left[\frac{1+s s'\cos{(\theta_{{\bm k},{\bm k}+{\bm q}})}}{2}\right]G^{(0)}_{s'}({\bm k}+{\bm q},i\omega_n+i\Omega_m)~,
\end{equation}
\end{widetext}
where $s=+$ for electron-doped systems and $s=-$ for 
hole-doped systems, $\beta=1/(k_{\rm B} T)$,
\begin{equation}\label{eq:ex+corr}
W(q,i\Omega)=V_q+V^2_q\chi_{nn}(q, i\Omega)~,
\end{equation}
\begin{equation}\label{eq:chi_RPA}
\chi_{nn}(q,i\Omega) =
\frac{\chi_0(q,i\Omega)}{1-V_q\chi_0(q,i\Omega)} \equiv \frac{\chi_0(q,i\Omega)}{\varepsilon(q,i\Omega)}
\end{equation} 
is the RPA density-density response function and $\varepsilon(q,i\Omega)$ is the RPA dielectric function.

In Eq.~(\ref{eq:sigma_rpa}) $\omega_n=(2n+1)\pi/\beta$ is a fermionic Matsubara frequency, while the sum runs over all the bosonic Matsubara frequencies $\Omega_m=2m\pi/\beta$. 
The first and second terms in Eq.~(\ref{eq:ex+corr}) are responsible for the exchange interaction between a quasiparticle and the occupied Fermi sea (including the negative energy component), 
and for the interaction with particle-hole and collective virtual fluctuations, respectively.  
The factor 
in square brackets in Eq.~(\ref{eq:sigma_rpa}), which depends on the angle $\theta_{{\bm k},{\bm k}+{\bm q}}$ between ${\bm k}$ and ${\bm k}+{\bm q}$, captures
the dependence of Coulomb scattering on the relative chirality $s s'$ of the interacting electrons. Finally, the Green's function
\begin{equation}\label{eq:greensfunction}
G^{(0)}_{s}({\bm k},i\omega) = \frac{1}{i\omega - \xi_s({\bf k})}
\end{equation}
describes the free propagation of states with wave vector ${\bm k}$, Dirac energy $ \xi_s({\bm k})=s \hbar v_{\rm F} k-\mu$ (relative to the chemical potential) and chirality $s=\pm 1$.

In the absence of the metal gate, this approach captures~\cite{polini_ssc_2007} 
the Gonz\'alez-Guinea-Vozmediano logarithmic behavior of the quasiparticle velocity $v^\star_{\rm F}$ at low densities~\cite{gonzalez_prb_1999}, while at the same time taking into account dynamical screening at the RPA level. 

Once again, the presence of a metal gate is here taken into account by employing the screened e-e interaction in Eq.~(\ref{eq:channelpotential}). 

\subsection{Analytical results at the exchange-only level}

Before turning to the presentation of our main numerical results, we would like to derive some useful analytical results for the TDOS to leading order in the limit $k_{\rm c} \gg k_{\rm F}, 1/d$ and at the exchange-only level.

We start by evaluating the exchange contribution to the ground-state energy. After simple algebraic manipulations on Eq.~(\ref{eq:exchange}) we arrive at the following result:
\begin{equation}\label{eq:exchange-explicit}
\delta \varepsilon_{\rm x}(n) = -\frac{\varepsilon_{\rm F}}{\pi}\alpha_{\rm ee}\int_0^{\Lambda}d{\bar q}~{\cal F}({\bar q}dk_{\rm F})~\ell({\bar q})~,
\end{equation}
where
\begin{equation}
\Lambda \equiv \frac{k_{\rm c}}{k_{\rm F}}
\end{equation}
is the ultraviolet cutoff measured in units of the Fermi wave number and 
\begin{equation}
\ell({\bar q}) \equiv \int_0^{+\infty}d{\bar \Omega}~\delta {\bar \chi}_0({\bar q}, i {\bar \Omega})
\end{equation}
with ${\bar q} = q/k_{\rm F}$ and ${\bar \Omega} = \hbar\Omega/\varepsilon_{\rm F}$. Finally, $\delta {\bar \chi}_0({\bar q}, i {\bar \Omega}) = \delta \chi_0({\bar q}, i {\bar \Omega})/N(\varepsilon_{\rm F})$ where $N(\varepsilon_{\rm F}) = N_{\rm f} \varepsilon_{\rm F}/(2\pi \hbar^2 v^2_{\rm F})$ is the density-of-states at the Fermi energy. 
It is possible to show~\cite{barlas_prl_2007} that
\begin{equation}\label{eq:Laurent}
\lim_{{\bar q}\to \infty}\ell({\bar q}) = -\frac{\pi}{6 {\bar q}} + {\cal O}(1/{\bar q}^{2})~.
\end{equation}

Let us first review the behavior of $\delta \varepsilon_{\rm x}(n)$ in the absence of a metal gate ($d \to \infty$). In this case the form factor ${\cal F}(x) = 1$. We find the following asymptotic behavior in the limit $\Lambda \gg 1$ ands in the absence of a metal gate~\cite{barlas_prl_2007}:
\begin{equation}
\delta \varepsilon_{\rm x}(n) = \frac{\alpha_{\rm ee}}{6}\ln(\Lambda) + {\rm regular~terms}~,
\end{equation}
where ``regular terms'' denotes terms that are {\it finite} in the limit $\Lambda \gg 1$. To this order of perturbation theory and to leading order in the limit $\Lambda \gg 1$ we therefore find
\begin{equation}\label{eq:K-leading}
\frac{K_0}{K} = 1+ \frac{\alpha_{\rm ee}}{4}\ln(\Lambda)~. 
\end{equation}
It is very well known~\cite{gonzalez_prb_1999,polini_ssc_2007,borghi_ssc_2009} that the leading-order asymptotic expansion for $\Lambda \gg1 $ on the right-hand side of Eq.~(\ref{eq:K-leading}) coincides with the expansion of the quasiparticle velocity. We are therefore led to conclude that, to first order in $e^2$ and to leading order in $\ln(\Lambda)$ in the limit $\Lambda \gg 1$, $K_0/K = v^\star_{\rm F}/v_{\rm F}$. This agrees with Eq.~(\ref{eq:Fermiliquid}) since $F^{\rm s}_0$ is zero to first order in electron-electron interactions. The suppression of the compressibility of the interacting system, i.e.~$K$, is tied to the quasiparticle velocity enhancement~\cite{barlas_prl_2007}.

In the presence of a metal gate, however, Eq.~(\ref{eq:K-leading}) may change. 
For sake of simplicity, we focus on the case $\epsilon_1 = \epsilon_2$. In this case ${\cal F}(x) = 1 -\exp(-2x)$. We therefore find
\begin{equation}\label{eq:logd}
\delta \varepsilon_{\rm x}(n) = -\frac{\varepsilon_{\rm F}}{\pi}\alpha_{\rm ee}\int_0^{\Lambda}d{\bar q}~[1 - \exp(-2{\bar q} \eta)]~\ell({\bar q})~,
\end{equation}
where we have introduced the dimensionless parameter $\eta = d k_{\rm F}$. Using Eq.~(\ref{eq:Laurent}) in Eq.~(\ref{eq:logd}) we find
\begin{eqnarray}\label{eq:calculation}
\lim_{k_{\rm c} \to \infty} \delta \varepsilon_{\rm x}(n) &=& \frac{\varepsilon_{\rm F}}{6}\alpha_{\rm ee}\int_{\Lambda_0}^{\Lambda}d{\bar q}~\frac{1 - \exp(-2{\bar q} \eta)}{\bar q}\nonumber\\
&=&\frac{\varepsilon_{\rm F}}{6}\alpha_{\rm ee}\int_{2\Lambda_0\eta}^{2dk_{\rm c}}dx~\frac{1 - e^{-x}}{x}~,
\end{eqnarray}
where we have introduced an infrared cutoff $\Lambda_0$, whose precise value is completely irrelevant to end of calculating the leading behavior of $\delta \varepsilon_{\rm x}(n)$ in the limit $k_{\rm c} \to \infty$. We now consider the case of a nearby gate, i.e.~the limit in which $\eta \to 0$ or, more explicitly, $d \ll 1/k_{\rm F}$. In this case, Eq.~(\ref{eq:calculation}) yields
\begin{equation}\label{eq:new_limit}
\delta \varepsilon_{\rm x}(n) =\frac{\varepsilon_{\rm F}}{6}\alpha_{\rm ee}\ln(d k_{\rm c}) + {\rm regular~terms}~,
\end{equation}
and, therefore,
\begin{equation}\label{eq:K-leading-metal-gate}
\frac{K_0}{K} = 1+ \frac{\alpha_{\rm ee}}{4}\ln(d k_{\rm c})~. 
\end{equation}
We note that, as expected, the suppression of the compressibility $K$ with respect to the non-interacting value $K_0$ in the case of a nearby gate ($dk_{\rm F} \ll 1$) is less severe than in the case of a distant gate ($dk_{\rm F} \gg 1$).

\section{Numerical results}
\label{sect:numericalresults}
\begin{figure}[t]
\begin{center}
\includegraphics[width=1.0\linewidth]{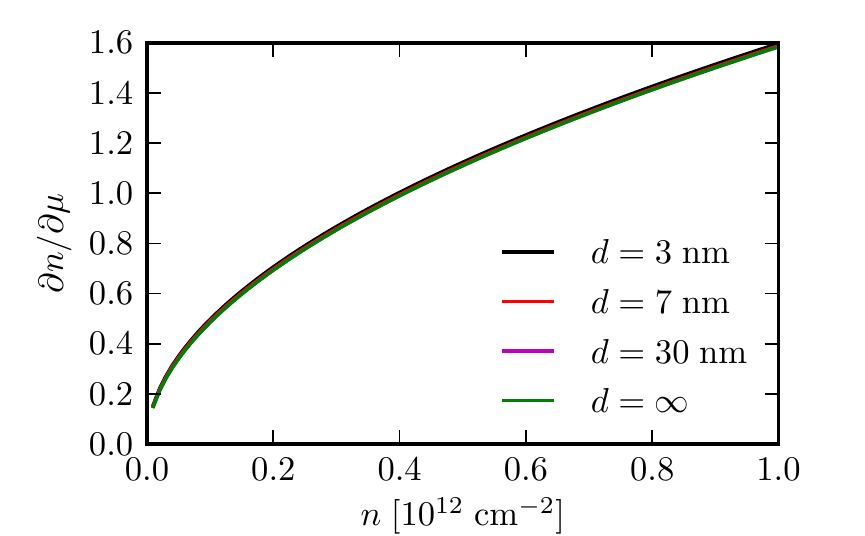}
\caption{(Color online) The TDOS $\partial n/\partial \mu$ (in units of ${\rm meV}^{-1}\times 10^{10} {\rm cm}^{-2}$) is plotted as a function of  carrier density $n$ in the range $1.0 \times 10^{10}~{\rm cm}^{-2} \leq n \leq 1.0 \times 10^{12}~{\rm cm}^{-2}$. Different curves correspond to different values of the graphene/metal gate distance $d$. The numerical results in this figure have been obtained by setting $\epsilon_1 = \epsilon_ 2 = 4.5$, corresponding to a graphene sheet encapsulated in hBN~\cite{yu_pnas_2013}. We note that in this case the dependence on the distance $d$ between graphene and the metal gate is negligible.\label{fig:two}}
\end{center}
\end{figure}

\begin{figure}[t]
\begin{center}
\includegraphics[width=1.0\linewidth]{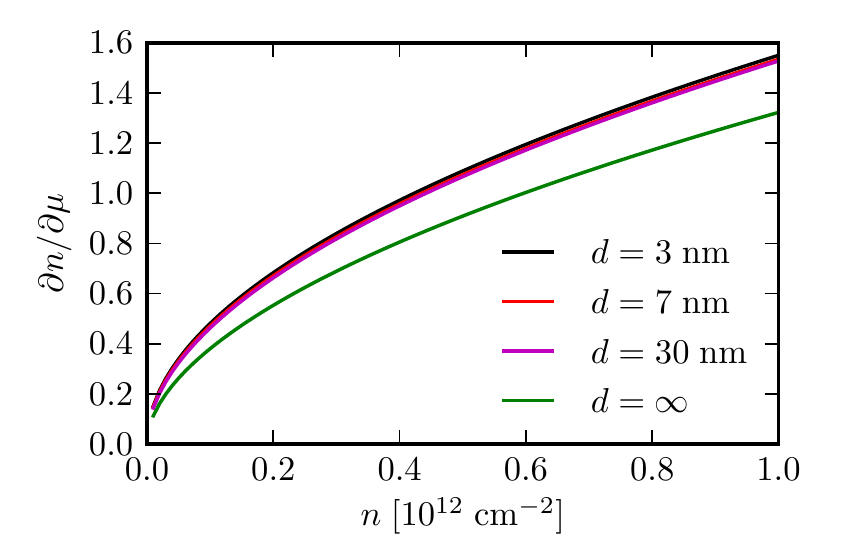}
\caption{(Color online) Same as in Fig.~\ref{fig:two} but for $\epsilon_1 = 4.5$ and $\epsilon_2 = 1.0$. In this case the TDOS in the presence of the metal gate is quite different from the one in the absence of a gate (curve labeled by $d = \infty$).\label{fig:three}}
\end{center}
\end{figure}

Figs.~\ref{fig:two}-\ref{fig:three} display our main numerical results for the TDOS of a graphene sheet in the presence of a metal gate.

In Fig.~\ref{fig:two} we illustrate the dependence of the TDOS $\partial n/\partial \mu$ on carrier density $n$ in the range $n = 1.0 \times 10^{10}~{\rm cm}^{-2}$ - $1.0 \times 10^{12}~{\rm cm}^{-2}$. This plot refers to the case $\epsilon_1 = \epsilon_2 = 4.5$. Different curves refer to different values of $d$. We see that in this case the TDOS displays a very weak dependence on $d$. The asymptotic result for a graphene sheet in the absence of a metal gate (curve labeled by $d = \infty$) is practically reached immediately. On the scale of the plot, results for a gate as close as $d = 3~{\rm nm}$ (roughly corresponding to $10$ hBN layers) are indistinguishable from the $d = \infty$ results.

The situation is rather different in the case $\epsilon_1 \neq \epsilon_2$. In this general case, the effective interaction (\ref{eq:channelpotential}) leads to a larger dependence of the TDOS on $d$. In Fig.~\ref{fig:three}, for example, we illustrate our predictions for $\partial n/\partial \mu$ in the case $\epsilon_1 = 4.5$ and $\epsilon_2 = 1$. In this case, even gates located as far as $30~{\rm nm}$  from the graphene sheet represent a severe obstacle in the quest of the asymptotic $d = \infty$ result.

\begin{figure}[t]
\begin{center}
\includegraphics[width=1.0\linewidth]{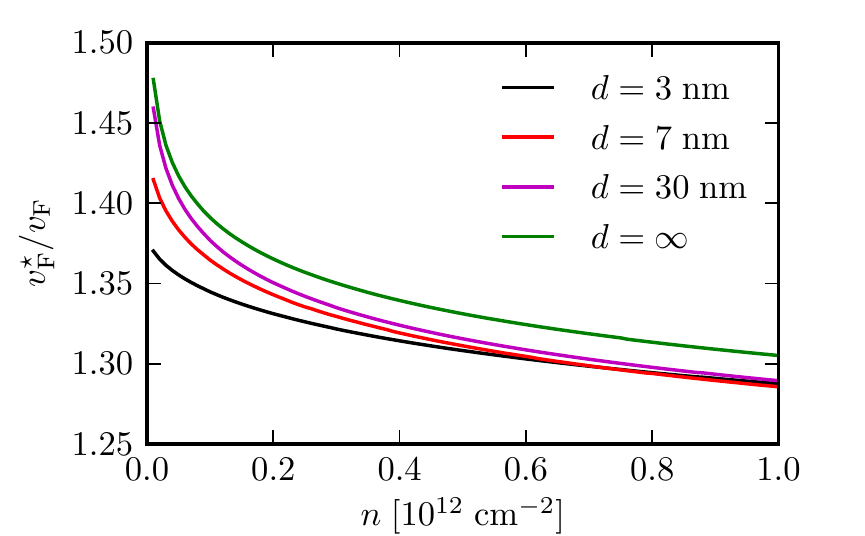}
\caption{(Color online) The quasiparticle velocity enhancement $v^\star_{\rm F}/v_{\rm F}$ is plotted as a function of carrier density $n$. As in Figs.~\ref{fig:two}-\ref{fig:three}, different curves refer to different values of the graphene/metal gate distance $d$. Data in this plot refer to the case~\cite{yu_pnas_2013} $\epsilon_1 = \epsilon_2 = 4.5$. As expected, decreasing $d$ for a fixed carrier density results in a decrease of the ratio $v^\star_{\rm F}/v_{\rm F}$.\label{fig:four}}
\end{center}
\end{figure}

We now turn to a brief discussion of the Landau parameter $F^{\rm s}_{0}$. This quantity can be easily accessed experimentally by measuring in the same setup {\it both} the TDOS $\partial n/\partial \mu$ and the quasiparticle velocity enhancement $v^\star_{\rm F}/v_{\rm F}$. Our numerical results for $v^\star_{\rm F}/v_{\rm F}$ are reported in Fig.~\ref{fig:four}. Data shown in this plot refer to the case $\epsilon_1 = \epsilon_2 = 4.5$. 

\begin{figure}[t]
\begin{center}
\includegraphics[width=1.0\linewidth]{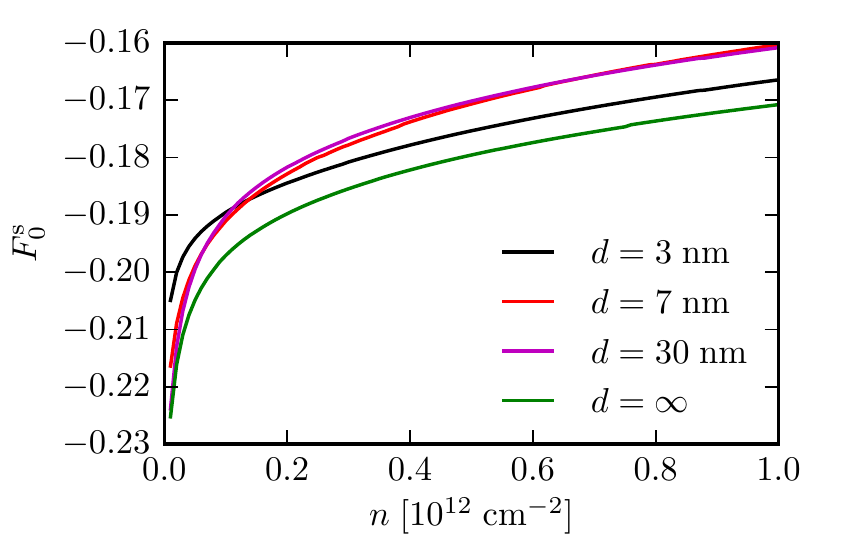}
\caption{(Color online) Theoretical predictions for the Landau parameter $F^{\rm s}_{0}$ of a 2D MDF fluid. The dimensionless quantity $F^{\rm s}_{0}$ calculated from Eq.~(\ref{eq:Fzeros}) is plotted as a function of carrier density $n$. Data in this plot refer to the case~\cite{yu_pnas_2013} $\epsilon_1 = \epsilon_2 = 4.5$.\label{fig:five}}
\end{center}
\end{figure}

Our predictions for $F^{\rm s}_{0}$ are shown in Fig.~\ref{fig:five}. These results have been obtained from 
\begin{equation}\label{eq:Fzeros}
F^{\rm s}_{0} =  \frac{v_{\rm F}/v^\star_{\rm F}}{K/K_0} -1~,
\end{equation}
which trivially descends from the Fermi-liquid formula (\ref{eq:Fermiliquid}).

\section{Summary and conclusions}
\label{sect:conclusions}
In summary, we have presented extensive numerical calculations based on the random phase approximation of the quantum capacitance and the spin- and circularly-symmetric Landau parameter of a two-dimensional fluid of massless Dirac fermions in a doped graphene sheet. 

With reference to recent experiments~\cite{yu_pnas_2013}, we have quantified the role of a metal gate, discovering that the quantum capacitance of a graphene sheet encapsulated between two media with identical (or similar) dielectric constants is nearly insensitive to the presence of the gate---see results in Fig.~\ref{fig:two}. 

Finally, we have pointed out that the combination of Shubnikov-de Haas transport experiments in a weak magnetic field {\it and} quantum capacitance measurements in the same sample allows to extract the value of the spin- and circularly-symmetric Landau parameter $F^{\rm s}_0$ of a two-dimensional fluid of massless Dirac fermions. These experiment may shed light on the role of vertex corrections in the many-body theory of two-dimensional massless Dirac fermion fluids. Our predictions for $F^{\rm s}_0$ are reported in Fig.~\ref{fig:five}.

In passing, we would like to remark that the thermodynamic density of states can also be calculated from an exact identity~\cite{katsnelson_prb_2000}, which can be easily derived from the sole use of the Luttinger theorem and Ward identities~\cite{Nozieres,Platzman_and_Wolff,shankar_rmp_1994,Giuliani_and_Vignale}. This is not the route we have followed here. The thermodynamic density of states calculated from the  derivative of the random-phase-approximation ground-state energy with respect to density does {\it not} coincide e.g.~with the one that can be calculated from the value of the retarded $G_0W$ self-energy (\ref{eq:sigma_rpa}) on the Fermi surface. We remind the reader that the quasiparticle self-energy has been used in this work to calculate the renormalized quasiparticle velocity $v^\star_{\rm F}$. This lack of ``internal consistency'' in the theory of the Fermi-liquid properties of 2D quantum electron liquids can be bypassed, at least in the charge channel where Luttinger theorem holds, by using the exact identities in Eqs.~(10)-(11) of Ref.~\onlinecite{katsnelson_prb_2000}.

\acknowledgements
This work was supported by the E.U. through the Graphene Flagship program  (contract no. CNECT-ICT-604391) (M.I.K. and M.P.) and the ERC Advanced Grant No. 338957 FEMTO/NANO (M.I.K.), and the Italian Ministry of Education, University, and Research (MIUR) through the programs ``FIRB - Futuro in Ricerca 2010" - Project PLASMOGRAPH (Grant No. RBFR10M5BT) and ``Progetti Premiali 2012" - Project ABNANOTECH (M.P.).

\end{document}